\documentclass[aps,pre,superscriptaddress]{revtex4}
\usepackage{graphicx}
\usepackage{amsmath}
\usepackage{amsfonts}
\usepackage{amssymb}
\usepackage{epsfig}
\usepackage{color}
\usepackage{bm}
\usepackage{wasysym}
\graphicspath{{pictures/}}
\begin{document}

\title{Powerful laser-produced quasi-half-cycle THz pulses}

\author{A. S. Kuratov}
\affiliation{P. N. Lebedev Physics Institute, Russian Academy of
Science, Leninskii Prospect 53, Moscow 119991, Russia}
\affiliation{Center for Fundamental and Applied Research,
Dukhov Research Institute of Automatics (VNIIA), Moscow 127055, Russia}
\author{A. V. Brantov}
\affiliation{P. N. Lebedev Physics Institute, Russian Academy of
Science, Leninskii Prospect 53, Moscow 119991, Russia}
\affiliation{Center for Fundamental and Applied Research,
Dukhov Research Institute of Automatics (VNIIA), Moscow 127055, Russia}
\author{V. F. Kovalev}
\affiliation{P. N. Lebedev Physics Institute, Russian Academy of
Science, Leninskii Prospect 53, Moscow 119991, Russia}
\affiliation{Keldysh Institute of Applied Mathematics, Russian Academy of Sciences, Moscow 125047, Russia}
\author{V. Yu. Bychenkov}
\affiliation{P. N. Lebedev Physics Institute, Russian Academy of
Science, Leninskii Prospect 53, Moscow 119991, Russia}
\affiliation{Center for Fundamental and Applied Research,
Dukhov Research Institute of Automatics (VNIIA), Moscow 127055, Russia}

\begin{abstract}
The Maxwell equations based 3D analytical solution for the terahertz half-cycle electromagnetic wave transition radiation pulse has been found. This solution describes generation and propagation of transition radiation into free space from laser-produced relativistic electron bunch crossing a target-vacuum interface as a result of ultrashort laser pulse interaction with a thin high-conductivity target. The analytical solution found complements the theory of laser initiated transition radiation by describing the generated THz wave shape at the arbitrary distance from the generating target surface domain including near-field zone rather than the standard far-field characterization. The analytical research has also been supplemented with the 3D simulations using the finite-difference time-domain (FDTD) method, which makes it possible for description of much wider spatial domain as compared to that from the particle-in-cell (PIC) approach. The results reported fundamentally shed light on the interfere of an electron bunch field and THz field of broadband transition radiation from laser-plasma interaction studied for a long time in the experiments with solid density plasma and may in future inspire them to targeted measurements and investigations of unique super intense half-cycle THz radiation waves near the laser target.
\end{abstract}

\maketitle

\section{Introduction}

Terahertz pulses with an ultrabroad spectral bandwidth \cite{liao20,gopal13,vicario,Shalaby,Tzortzakis} attract a lot of attention due to the large number of applications in various fields of science and technology~\cite{salen,Weightman,Kampfrath,Minty}. The maximum energy yield of a THz pulse can be achieved in the interaction of a short intense laser pulse with overdense plasma \cite{liao20} that pushes the boundaries of possible applications of THz radiation to new areas, which require high energies of generated pulses~\cite{salen}. The broadest spectral bandwidth is achieved when a half-cycle (unipolar) THz pulse is generated, that has already been known because charged particles can radiate in the form of unipolar pulses (see review \cite{Rosanov} and Refs. therein). Emission in such a form is still often considered as a "strange electromagnetic wave" \cite{Vinogradov}, since it looks like DC field simply propagating with speed of light, but satisfies Maxwell's equations in vacuum as a standard electromagnetic wave. To our best knowledge, such half-cycle type (quasi-half-cycle) pulses generated in laser-target interactions were discussed only in numerical simulations or proposed as {\it ad hoc} a guess \cite{Ding, liao19}, while quasi-half-cycle pulses have already been observed in the experiments with electron beams from conventional accelerators \cite{Wu13}. Standardly, theory of transition radiation from electron bunch is presented in the far-field approximation \cite{tilborg04,root07,tr1,tr2} as extension of classical theory for a single electron \cite{Ginzburg1946}. There is no corresponding near-field theory, which could be a base for the experimental measurements THz fields near the target surface, where these fields have maximum strength. The problem is that special detectors are needed to respond to fast static traveling field in free space. This is one of the reasons for low motivation to targeted experiments on namely high-intense unipolar radiation THz pulses, whose field profile measurement still faces technical difficulties. At the same time, it was predicted that such powerful THz pulses could provide effective control over molecules adsorbed to surfaces and ferroelectric polarization and molecular orientation \cite{salen}.

Possible applications and development of appropriate diagnostics, e.g., with advanced  electro-optic detection \cite{eod} and electron/proton radiography \cite{Quinn,Inoue}, together with adequate theoretical support could be a good background in pursuing the works on laser-produced powerful half-cycle pulse source. For example, highly beneficial novel THz light sources could be those, which reach the spectral range from few to 15 THz, where the present laser THz sources based on optical rectification \cite{vicario} usually do not operate because of strong absorption in crystals. Conventional designs of accelerator-based sources face difficulties in obtaining too short electron bunches necessary to achieve these frequencies. However, this is easily overcome by using electron bunches generated in interaction of femtosecond laser pulses with a plasma \cite{Inoue}. The highest density of electron current can be obtained in interaction with solid dense target and here we focus on such laser-target design for generation of terahertz radiation unipolar pulses of extreme frequency range (up to 15 THz) by femtosecond lasers. Note, that laser production of extremely intense THz pulses could be state-of-art work in the context of terahertz-driven electron acceleration \cite{Nanni}.

Mechanisms of THz generation in short intense laser pulse interaction with overdense plasma are associated with effective heating and ejection of target electrons and widely discussed \cite{consoli20,liao20}. Most of the laser heated electrons are trapped in a sheath field layer at the target-vacuum interface. A sheath plasma expansion model has been developed to describe THz radiation of $\sim 1$ THz frequency in the direction perpendicular to the laser pulse propagation direction \cite{liao20,gopal13,gopalol,gopal}. Transition radiation of laser heated electrons leaving the target is considered as another typical mechanism of THz generation \cite{tilborg04,root07,tr1,tr2}. For the highly relativistic electron beam this mechanism generates well collimated THz pulses along electron propagation direction. The electrons trapped in the sheath experience both deceleration/acceleration and reversals at the spatial scale of the hot electron Debye length producing Bremsstrahlung and synchrotron radiation, correspondingly. These radiation mechanisms have efficiencies comparable to transition radiation, but belongs to an optical range with typical frequency close on the order of magnitude to the laser frequency, i.e. one may refer them to re-emission of the laser light by laser-accelerated electrons. Focusing on highly beneficial novel light sources of extreme frequency range (up to 15 THz) based on femtosecond lasers, here we consider a transition radiation mechanism.

Until now all the measurements of THz pulses were performed in the far-field zone and standartly are based on only theoretical description of the asymptotic characteristics of wave energy and angular-spectral distributions. However, the nature of generated THz pulse can be comprehended only through rigorous analytical wave theory without far-field approximation. Neither such experimental measurements nor far-field theory are able to reconstruct a THz pulse shape yet. The theory presented fills this gap. We present an analytical solution to Maxwell equations in the near-field zone for an ideal conductivity target and compare it with a result of numerical model for the case of a finite target conductivity. Our analytical model (albeit with some simplifications for the sake of analytical calculations) clearly demonstrates that the coherent transition radiation of an electron bunch at the target-vacuum interface has the form of a half-cycle terahertz pulse. Certainly, there is a great advantage of a theory over numerical simulations (see, e.g. \cite{Ding}) since its ability to explicitly scale THz field characteristics vs laser parameters.

\section{Analytical solution}\label{sec:examples}

Starting with Maxwell's equations, consider the process of electromagnetic field generation into a vacuum ($z>0$) by laser-induced electron current, ${\bf j} = (0,0,j_z)$, injected from an ideal conductor ($z<0$), e.g. from a high-conductivity plasma back side. This is a given source in Maxwell's equations in the form of electrical current of most energetic electrons, which are accelerated by a laser pulse in the forward direction, along the Z axis, and have an energy enough to overcome a sheath potential barrier. Let, for definiteness, this current appears at $t=0$ and the plasma-vacuum interface has an ideal interface (semi-bounded plasma). In fact, the latter assumption is valid as long as the size of the sometimes possible preplasma is less than the wavelengthh of the generated electromagnetic wave of our interest (THz range). We consider infinite boundary between target and vacuum, that is good approximation as long as $b=\pi L/\lambda \gamma \gg 1$, where $L$ is the transverse size of the target boundary, $\lambda$ is the characteristic wavelength of the radiated wave, and $\gamma$ is the electron beam gamma factor. The latter makes it possible to neglect the contribution of diffraction radiation, which, of course, produces wings of opposite polarity in the profile of the generated field, but introduces an error that is exponentially small $\sim \exp{(-b)}$ \cite{tilborg04}.

Given a single nonzero component of the electric current ($j_z \neq 0$), the emitted into a vacuum electromagnetic field is characterized by the following electric, ${\bf E} = (E_{\rho},0,E_z)$, and  magnetic, ${\bf H} = (0,H_{\varphi},0)$, components. By introducing the vector (${\bf A} = (0,0,A_z)$) and scalar ($\phi$) potentials, ${\bf H} = {\rm rot} {\bf A}$ and ${\bf E} = - \nabla \phi - (1/c) \partial_t {\bf A}$, which obey the Lorentz gauge condition, $\partial_z A_z + \partial_t \phi /c =0$, the Maxwell's equations are reduced to a single equation for $A_z$
\begin{equation}
\partial_{tt} A_z = c^2 \triangle A_z + 4 \pi c j_z \,. \label{vecpoteqn}
\end{equation}
The solution of (\ref{vecpoteqn}) in a free space reads: $A_z (t, {\bf r}) = \int d^3 {\bf r}' j_z (t-\eta/c, {\bf r}')/(\eta c)$, where $\eta = |{\bf r} -{\bf r}'|$ and integration is over the domain $\eta < c t $. We use the cylindrical symmetry and the factorized form of the electrical current $j_z = Q v n_z(t, z) n_\perp(\rho)$, where $Q$ is the total charge of the electron bunch moving with the velocity $v=const$, $n_z$ is the linear electron density distribution versus $z$ and $t$, and $n_\perp$ is the areal electron density distribution versus $\rho =\sqrt{x^2+y^2}$. Here, a simplifying assumption of a given constant electron velocity is used in order to explicitly obtain an analytical solution for the generated transition radiation electromagnetic pulse. Then, the solution of (\ref{vecpoteqn}) can be written as follows
\begin{equation}
A_z = \frac{Q v}{c} \!\!\!\!\!\!\!\!\!\!\!\!  \int \limits_{\sqrt{\rho'^2+z'^2} < c t} \!\!\!\!\!\! \!\! \!\! \!\! {\rm d} z' {\rm d} \rho' \rho' N_\perp(\rho,\rho') \frac{n_z (t-\sqrt{\rho'^2+z'^2}/c, z'  + z) } {\sqrt{\rho'^2+z'^2}} \,,
\label{az0}
\end{equation}
where
\begin{equation}
		N_{\perp}(\rho,\rho') = \int\limits_{0}^{2\pi} {\rm d}\chi \,   n_{\perp} \left( \sqrt{\rho^2 + \rho'^2 + 2\rho' \rho \cos \chi} \right)  \,.
	\label{dens_transverse}
\end{equation}
The desired solution should meet the boundary condition of vanishing of the tangential electric field component $E_{\rho}$ at the vacuum-target interface, $E_{\rho}\mid_{z=0} = 0 $. This is achieved using the so called image method when the desired electromagnetic field can be represented as a superposition of two free space type fields given by (\ref{az0}). They are generated by two currents, $j_z^+$ and $j_z^-$, having the charges of opposite signs and moving from the vacuum-target interface in two opposite directions. Correspondingly, $j_z^+\equiv j_z$ with $n_z^+\equiv n_z(t,z)$ and  $j_z^-(t,\rho,z) \equiv j_z(t,\rho,-z)$ with $n_z^- \equiv n_z(t,-z) $ and the substitution of $n_z^{\pm}$ into (\ref{az0}) makes it possible to obtain in explicit form $z$-components of the vector potential, $A_z^+$ and $A_z^-$, and hence the desired solution for the electromagnetic field in a half-space $z>0$

\begin{eqnarray} \label{eb}
&& E_z = - (1/c) \, \partial_{t} (A_z^{+}+A_z^{-})
      - (c/v) \partial_{z} (A_z^{+}-A_z^{-}) \,, \\ \nonumber  &&
E_\rho = - (c/v) \partial_{\rho} (A_z^{+}-A_z^{-}) \,, \quad
H_\varphi = - \partial_{\rho} (A_z^{+}+A_z^{-}) \,,
\end{eqnarray}
where it was used that $\phi^+ + \phi^- = (c/v) (A_z^+ - A_z^-)$.

To concretize the solution (\ref{az0}) we specify spatial-temporal form of the electron bunch by introducing $n_z^\pm = \theta(\pm z)(\theta(v t\mp z)-\theta(v \tau \mp z))/(v t_0)$, where $\tau = t-t_0$, and $n_\perp = \exp \left(- \rho^2/r_0^2 \right)/(\pi r_0^2)$. Here the Heaviside step function, $\theta(t)$, corresponds to the simplified rectangular time shape of the electron bunch with the duration $t_0$ and the transverse Gaussian distribution with the characteristic radius $r_0$. The latter can be addressed to a laser pulse with approximately the same duration and spot size radius $\lesssim r_0$. Given this specification one can write $N_\perp$ in the following explicit form
\begin{equation}
N_{\perp}^{gs}(\rho,\rho') = \frac{2}{r_0^2} \, I_0 \left( \frac{2 \rho' \rho}{r_0^2}\right) \exp \left( - \frac{\rho^2+\rho'^2}{r_0^2} \right)  \,,
\label{dens_gauss}
\end{equation}
where $I_0$ is the modified Bessel function, and reduce $A_z^\pm$ to a simple integral form
\begin{eqnarray}
& A_z^{\pm} =  \frac{Q}{c t_0} \, \int\limits_{0}^{\infty} \rho' {\rm d}\rho' N_{\perp}^{gs}(\rho,\rho') [F^{\pm}(t)-F^{\pm}(\tau)] \,, \label{az} \\
& F^{\pm} = \theta \left( \sqrt{c^2 t^2-z^2}-\rho'\right)  \ln
\left( \cfrac{v t \mp z + R^\pm (\rho') }{(1+\beta)(\sqrt{z^2+\rho'^2}\mp z)}\right) \, , \nonumber
\end{eqnarray}
where $\beta = v/c$ and $R^\pm (\rho') =\sqrt{(v t \mp z)^2 + (1-\beta^2)\rho'^2}$. The using \eqref{az} in \eqref{eb} makes it possible to analyze in detail the structure of the generated electromagnetic field.

\subsection{The case of the sausage type electron bunch}

The most intense femtosecond laser pulse, which is able to produce the highest electron current density and, hence, the most powerful THz pulse should be focused into the few micron focal spot, that typically corresponds to $c t_0 \gg r_0$. In this case, the transversal Gaussian electron beam profile can be replaced by the delta-functional distribution, $N_{\perp}^p(\rho,\rho') = \delta(\rho-\rho')/\rho $ in Eq. \eqref{az}. Correspondingly, from  Eq. \eqref{az} we arrive to the following easy-to-analyzee analytical expressions for the electromagnetic field components


\begin{eqnarray} \label{line}
\nonumber
&& E_z = \frac{Q}{v t_0} \left[ \left(\frac{1-\beta^2}{R^+(t)} + \frac{1-\beta^2}{R^-(t)} - \frac{2}{r} \right) \theta(c t - r) -  \left(\frac{1-\beta^2}{R^+(\tau)} + \frac{1-\beta^2}{R^-(\tau)} - \frac{2}{r} \right) \theta(c \tau - r) \right] \,, \\
&& E_\rho = \frac{Q} {v t_0 \rho} \left[ \left(\frac{v t -z}{R^+(t)} - \frac{vt +z}{R^-(t)} + \frac{2 z}{r} \right) \theta(c t - r) -  \left(\frac{v \tau -z}{R^+(\tau)} - \frac{v \tau+z}{R^-(\tau)} +\frac{2 z}{r} \right) \theta(c \tau - r) \right] \,, \\ \nonumber
&& H_\varphi = \frac{Q}{c t_0 \rho} \left[ \left(\frac{v t -z}{R^+(t)} + \frac{vt +z}{R^-(t)} \right) \theta(c t - r) -  \left(\frac{v \tau -z}{R^+(\tau)} + \frac{v \tau+z}{R^-(\tau)} \right) \theta(c \tau - r) \right] \,,
\end{eqnarray}
where $r = \sqrt{z^2 + \rho^2}$. In the limit $t_0 \rightarrow 0$ from Eqs. \eqref{line} we recover the known result for a point charge \cite{bolotovskii}. \par

In the general case, Eqs. \eqref{line} do not demonstrate a simple separation of the fields, the intrinsic  bunch field and the radiation field.  In the far-field zone, $ c t > r \gg c t_0$ the electric field components (\ref{line}) can be rewritten in the following form
 \begin{equation}
 	\label{fzone}
 \begin{aligned}
	E_z & = \frac{Q}{\beta } \frac{\theta(c t - r) - \theta(c \tau - r)}{c t_0}  \left(\frac{1-\beta^2}{R^+(t)} + \frac{1-\beta^2}{R^-(t)} - \frac{2}{r} \right) \\
	&
	- Q (1-\beta^2) \theta(c \tau - r)
	  \left(\frac{v t -z}{(R^+(\tau))^3} + \frac{v t +z}{(R^-(\tau))^3} \right) \, ,  \\
	E_\rho & = \frac{Q} {\beta \rho}
	\frac{\theta(c t - r) - \theta(c \tau - r)}{c t_0}
	\left(\frac{v t -z}{R^+(t)} - \frac{vt +z}{R^-(t)} + \frac{2 z}{r} \right) 	
	 \\ &
	 	+ Q \rho (1-\beta^2) \theta(c \tau - r)
	 \left(\frac{1}{(R^+(\tau))^3} - \frac{1}{(R^-(\tau))^3} \right) \, ,  \\
	H_\varphi & = \frac{Q} {\rho}
	\frac{\theta(c t - r) - \theta(c \tau - r)}{c t_0}
	\left(\frac{v t -z}{R^+(t)} + \frac{vt +z}{R^-(t)} \right) 	
	\\ &
	+ Q \rho \beta (1-\beta^2) \theta(c \tau - r)
	\left(\frac{1}{(R^+(\tau))^3} + \frac{1}{(R^-(\tau))^3} \right) \, ,
\end{aligned}
\end{equation}
where we have neglected all the terms decreasing faster than $1/r^2$, and denoted $\rho = r \sin \vartheta$ and $z =r \cos \vartheta$.

Each of the electromagnetic field components in (\ref{fzone}) has two distinct contributions: $\textbf{E} = \textbf{E}^{rad} + \textbf{E}^{int}$ and $\textbf{H} = \textbf{H}^{rad} + \textbf{H}^{int}$. The contributions $\textbf{E}^{rad}$ and $\textbf{H}^{rad}$ proportional to the difference of $\theta$-functions decrease as $1/r$ for large $r$ and define the radiation field, while the remaining contributions $\textbf{E}^{int}$ and $\textbf{H}^{int}$ decreasing as $1/r^2$ for large $r$ describe the intrinsic  field of the moving electron bunch. Propagating radiation field reaches the given point at the distance $r$ in the far-field zone at the instant $t=r/c$ and lasts till $t = r/c + t_0$. Then, as times goes by, a radiation field is replaced by a weak incoming intrinsic field (see, for example, Fig.\ref{figH0} below). For the instant corresponding to onset of a radiation field, $ct \simeq r$, one arrives to a simple form for the radiation components, $E_r = E_\rho \sin \vartheta + E_z \cos \vartheta = 0$ and $E_\vartheta = E_\rho \cos \vartheta - E_z \sin \vartheta = H_\phi$, which can be rewritten for the better clarity in the spherical coordinates as following
\begin{equation}
	\label{far_zone_sphere}
		E_{\vartheta}^{rad} = H_{\varphi}^{rad} \simeq \frac{2 Q }{r} \frac{\beta \sin\vartheta}{ 1 - \beta^2\cos^2 \vartheta} \frac{\theta(c t - r) - \theta(c \tau - r)}{c t_0}
        \,, \quad E_r^{rad} \simeq 0 \,.
\end{equation}
As expected, these results show, that the far-zone radiation field is a spherical transverse electromagnetic wave with the amplitude decreasing  $\propto 1/r$.
And again for explicitness, with extreme particle bunch shortening $ c t_0 \to 0 $ the difference between two $\theta$-functions in (\ref{far_zone_sphere}) can be replaced by the $\delta$-function,  $(\theta(c t - r) - \theta(c \tau - r))/(c t_0) \to \delta (ct-r) $, and we arrive to the formula for the transition radiation field generated by a point charge \cite{bolotovskii}. \par

\begin{figure} [!ht]
\centering \includegraphics[width=0.4 \linewidth]{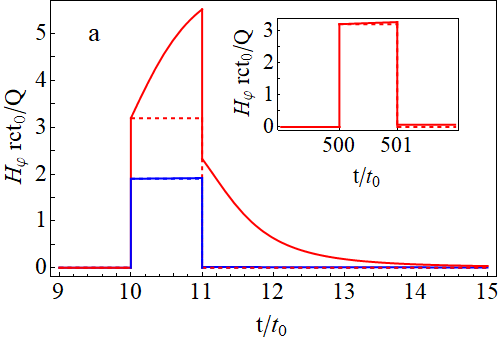}
\includegraphics[width=0.4 \linewidth]{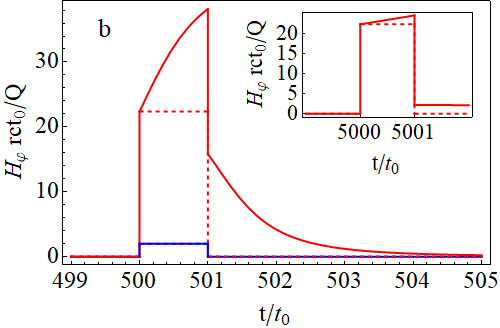}
\caption{Magnetic field temporal profile, $H_{\varphi}(t)$, for $v=0.95 \,c$ ($\gamma \simeq 3.2$) at $r= 10 \,c t_0$ (left) and for $v=0.999 \,c$ ($\gamma \simeq 22$) at $r= 500 \,c t_0$ (right) along the direction, $\vartheta = 1/\gamma$, of maximum radiated field (red curves) and along the target surface, $\vartheta = \pi/2$, (blue curves). The dashed lines corresponds to the far-field approximation. The insets show magnetic field temporal profiles at $ r= 500\, c t_0$ (left) and $ r= 5000\, c t_0$ (right).} \label{figH0}
\end{figure}

The electromagnetic field temporal profile has a form of half-cycle pulse (see Fig. \ref{figH0}) with the width defined by the electron pulse duration, $ t_0$. For the given total electron charge the field amplitude increases with energy of the electron beam. For the ultrarelativistic electrons with $\gamma = 1/\sqrt{1-\beta^2}\gg 1$ the direction of radiated field maximum corresponds to $\theta \simeq 1/\gamma $. In this direction, separation of the generated electromagnetic field into the intrinsic field of the moving charge and the radiated field is not possible in the near-field zone. However, this is possible at the large distances, $r\ggg ct_0$, where a far-field approximation works and radiation field amplitude drops significantly (see insets in Fig. \ref{figH0}). Selection of the radiation component from the total electromagnetic field in the direction of its maximum intensity is complicated for very energetic electrons. The higher their energy the longer distance is required to measure a true field of radiation. For example, an optimal angle of radiation of 22 MeV electrons ($v\simeq 0.999 c$) is only $\sim 2.5^\circ$ and therefore intrinsic field will have negligible contribution only at the distance longer than 5000 $c t_0$, i.e. $\sim 15$ cm for $t_0 = 100$ fs (see Fig. \ref{figH0}). On the other hand, in the transverse direction such selection is possible at much shorter distances in accordance with Fig. \ref{figH0}, where there is no visible difference between blue dashed and solid curves. Thus, the finding of the true radiation field energy may require corresponding recalculating it from the measured total field energy by taking into account the theoretical space-angular finding presented above.

\begin{figure} [!ht]
\centering \includegraphics[width=0.4 \linewidth]{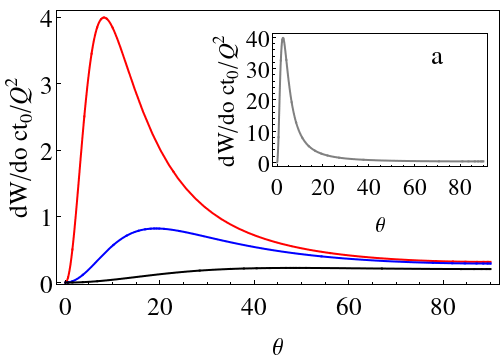}
\includegraphics[width=0.4 \linewidth]{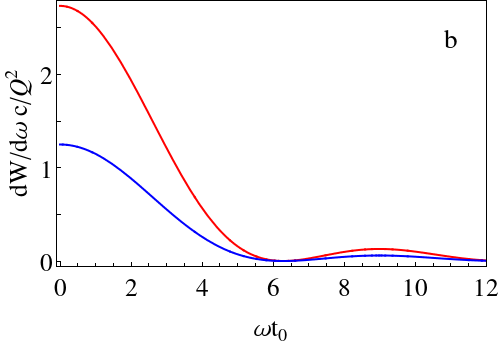}
\caption{Angular distribution of radiated energy (left) for $v=0.99 c$ (red), $v=0.95 c$ (blue), $v=0.8 c$ (black) and $v=0.999 c$ (gray in inset). Spectra of radiated energy (right) for $v=0.99 c$ (red) and $v=0.9 c$ (blue).} \label{figW}
\end{figure}

The spectral-angular distribution of radiated energy in a far-field zone is
\begin{equation} \label{eqW}
 \frac{d W}{d o d \omega} = \frac{c r^2}{4 \pi^2} |H_\omega|^2 = \frac{4Q^2}{\pi^2 c} \frac{\beta^2 \sin^2 \vartheta}{(1-\beta^2 \cos^2 \vartheta)^2} \left | \frac{\sin (\omega t_0/2)}{\omega t_0} \right|^2\,.
\end{equation}
It demonstrates a classical angular distribution \cite{ginzburg} with maximum at the angle $\theta \sim 1/\gamma $ for relativistic limit (see Fig. \ref{figW}), radiated energy decrease with frequency, and the spectrum width at half maximum $\Delta\omega_c \simeq 2.8/t_0 $ (see Fig. \ref{figW}). The latter naturally corresponds to the range, where the coherent transition radiation occurs, $\omega < t_0^{-1}$, and small incoherent contribution relevant to the higher frequencies. Oscillations in the high-frequency tail of the spectrum reflect only the model rectangular shape of the laser pulse, adopted for simplicity in order to achieve maximum clarity of the analytical description, and may not appear in the case of a natural smooth pulse. For ultrarelativistic electrons with $\gamma \gg 1$ the total radiated energy is well approximated by following simple expression  $W = Q^2 (4 \ln 2 \gamma -2 )/c t_0$.

The simplest reproduction of the above analytical result in the far-field zone and nonrelativistic case, $\beta \ll 1$, can be easily done in the dipole approximation. In this limits, by using $n_\perp = \delta(\rho-\rho')/(2 \pi \rho)$ in the density electron current and ${\bf H} = [\dot{\bf A} \times {\bf n}]/c $, where ${\bf n}$ is the unit vector along the radiation propagation direction and $A$ is given at the retarded time $t -r/c$ one gets
\begin{equation}\label{eqdip}
{\bf A} = \frac{1}{c r} \int {\bf j} d V = \frac{Q {\bf v}}{c r} \left(\int \limits_0^\infty n_z^+ d z - \int \limits_0^{-\infty} n_z^- d z \right ) \,, \quad \dot{\bf A} =  \frac{2 Q {\bf v}}{c r t_0}(\theta(t') - \theta(t'-t_0))\,\
\end{equation}
Then, Eqs. \eqref{eqdip} lead to the field components coinciding with Eq. \eqref{far_zone_sphere} at $\beta \ll 1$.
The key point is that the time change of the dipole moment occurs not due to the particle velocity change (nonzero acceleration) rather than due to the bunch charge change, which increases as the bunch exits into a vacuum while being zero inside a target of high-conductivity.

\subsection{The case of arbitrary longitudinal and transverse widths of the bunch}\label{secB}

\begin{figure} [!ht]
\centering \includegraphics[width=0.4 \linewidth]{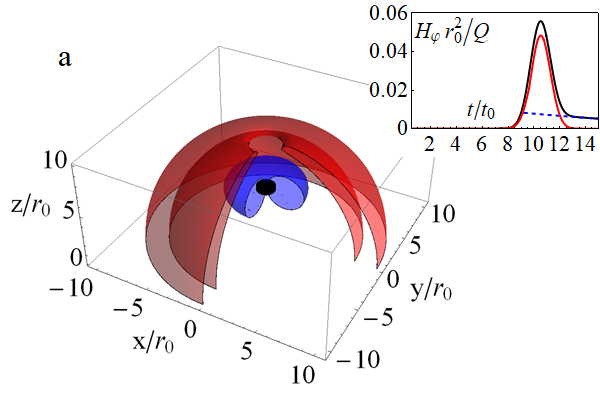}
\includegraphics[width=0.4 \linewidth]{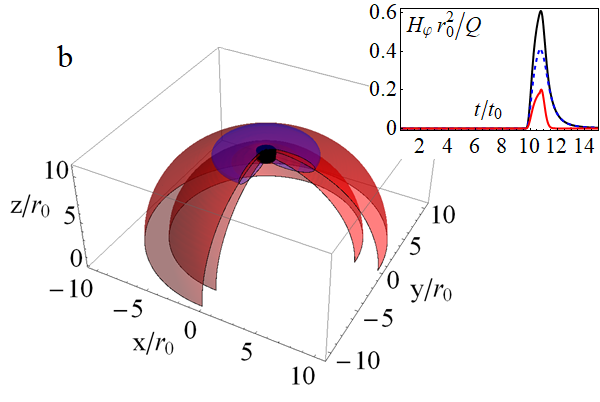}
\caption{Magnetic component $H_{\phi}$ of the generated electromagnetic field propagating in a vacuum for $v=0.5\, c$ (a) and for $v=0.95\,c$ (b) at the instant $t=10 t_0$. Electron bunch with the sizes $c t_0 = r_0$ is shown in black. The blue corresponding to the levels 0.05  $Q/r_0^2$ (a) and 0.1 $Q/r_0^2$ (b) show the intrinsic field, while the red ones for the levels 0.02  $Q/r_0^2$ (a) and  0.04 $Q/r_0^2$ (b), illustrate a radiated field.
The insets show the magnetic pulse time shape (black curve -- total field, blue dashed curve -- intrinsic field and red curve -- radiation field) at the distance of $10 r_0$ along the direction of maximum radiated field. } \label{figB}
\end{figure}

Let now turn to the general case described by Eqs. \eqref{eb}, \eqref{dens_gauss}, \eqref{az}. Standardly generated electromagnetic field (\ref{eb}) has two contributions, (1) the intrinsic field of a moving charge and of its image and (2) the radiation field. Such field structure is illustrated by Fig. \ref{figB}, where the intrinsic field of the moving charge is shown in blue and the radiation field in red. Formation of the spherical wave is clearly seen as well as the unipolar pulse of the radiation field (see insets in Fig. \ref{figB}).
The wave temporal profiles for the inserts in Fig. \ref{figB}a and Fig. \ref{figB}b are presented for different propagation directions, that results in a different relation between the intrinsic and radiation fields in these inserts. For the case $v=0.5\, c$ the temporal pulse was detected in the direction along a target-vacuum interface far enough from the electron bunch, while for the case $v=0.95\, c$ the field impulse was registered at small ($20^{\rm o}$) angular deviation from the bunch propagation direction. It is clearly seen that presented in Fig. \ref{figB} theoretically derived field structures qualitatively  correspond to the numerical simulation results on THz emission in forward direction from irradiated foil \cite{Ding}.

\begin{figure} [!ht]
\centering \includegraphics[width=0.4 \linewidth]{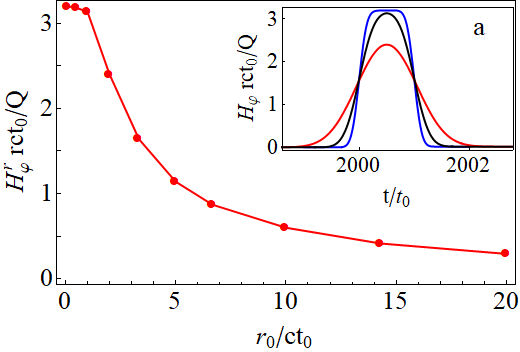}
\includegraphics[width=0.4 \linewidth]{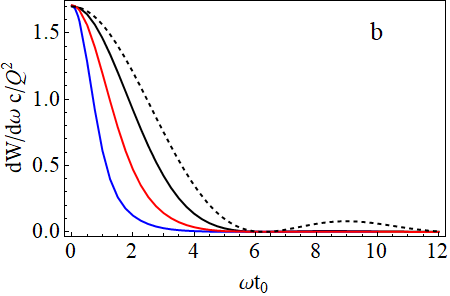}
\caption{The radiated magnetic field maximum vs ratio $r_0/ c t_0$ and spectra of radiated energy for $r_0 = 0.5 c t_0$ (black curves), $r_0 = c t_0$ (red curves) and $r_0 = 2 c t_0$ (blue curves) for $\beta = 0.95$. Dashed black curves corresponds to limiting case of $c t_0 \gg r_0$ (Eq. \eqref{eqW}). The insets show the magnetic pulse time shape at the distance of $2000 c t_0$ along the direction of maximum radiated field.} \label{figB1}
\end{figure}

With the broadening of the electron beam diameter, the profile of the generated pulse is smoothed out as soon as the transverse size of the beam approaches the longitudinal one. For $r_0 \sim ct_0$ the temporal field pulse profile takes on a Gaussian temporal shape relevant to the spatial Gaussian distribution of the electron beam (see inset in Fig.\ref{figB1}). For a given electron bunch charge (given laser power, see below) the higher the ration $c t_0/r_0$, the higher the electromagnetic pulse amplitude (see Fig.\ref{figB1}). If $r_0 \ll ct_0$ radiated pulse temporal shape follows the electron bunch temporal profile, e.g. rectangular one above discussed (see insets in Fig. \ref{figH0}). For a given $t_0$ the maximum filed decreases as $\sim 1/r_0$  as illustrated in Fig.\ref{figB1}a.

For relatively low bunch velocity $v \sim 0.5 c$ the radiation propagates predominantly along the target surface, while for the ultra-relativistic electrons a radiation pulse collimates along electron bunch propagation direction slightly shifting from it in accordance with a classical theory of transition radiation \cite{ginzburg, QE16}. The higher velocity the smaller this shift is.

A radiation spectrum is defined by the electron bunch spatial-temporal shape. The half-cycle THz field profile is clearly illustrated by well pronounced low-frequency spectrum domain, where a cutoff of the spectrum may appear in the case $b\sim 1$ due to the diffraction radiation contribution. Super-broadband emission ( Fig. \ref{figW}b) is characterized by the spectral bandwidth $\Delta\omega \simeq c/(c t_0 + r_0)$ in agreement with far-field approach \cite{QE16}. We note the dependence of the spectrum width on the focal spot size (electron bunch radius). Spectrum shrinks with increasing hot spot size. Correspondingly, total emitted energy  decreases with this size.
In the case of $\gamma t_0/(c r_0) \gg 1$ the total radiated energy can be estimated as $W_R = Q^2/(\pi c t_0) (3 \ln({\cal{E}}_e t_0/(m c r_0))-1 )$  \cite{QE16}, where ${\cal{E}}_e = m c^2\gamma $ is the  energy of the  laser-heated electron.

\section{The FDTD simulations}

To study effect of high but finite target conductivity on terahertz pulse generation we performed the 3D simulations with FDTD (finite-difference time-domain) method based on numerical solution of the Maxwell's equations  in a medium with a given dielectric susceptibility. We applied this simulation to a metal
target, where the dielectric permittivity is a complex function
($\epsilon\prime+i\epsilon\prime\prime$) but still $|\epsilon|\gg1$.
To describe a dielectric permittivity we used the standard Drude model $\epsilon = 1 + 4 \pi \sigma (\omega)/\omega$ with the conductivity $\sigma = \sigma_0/(1 - i \omega)/\nu)$, where $\sigma_0 = 10^{18}$~s$^{-1}$ and $\nu = 10^{13}$~s$^{-1}$.
The target occupied a
half-space $z<0\,\mu$m in the simulation box $-300\,\mu$m$<x<300\,\mu$m,
$-300\,\mu$m$<y<300\,\mu$m, and $-200\,\mu$m$<z<400\,\mu$m. The grid cell
size was $1\,\mu$m and the time step was 1\,fs. The distributed charge has the Gaussian profiles in both transverse directions, $z$ and $\rho = \sqrt{x^2+y^2}$, with the same hot spot size as above to compare with the theory developed, $c t_0=r_0 = 20 \,\mu$m. The bunch starts to move from the target surface along the normal, along the Z-axis, with a given velocity $ v = 0.5 c $ or $ v = 0.95 c $.

\begin{figure} [!ht]
\centering \includegraphics[width=0.35 \linewidth]{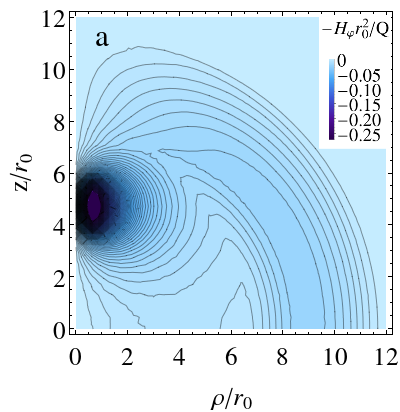}
\includegraphics[width=0.35 \linewidth]{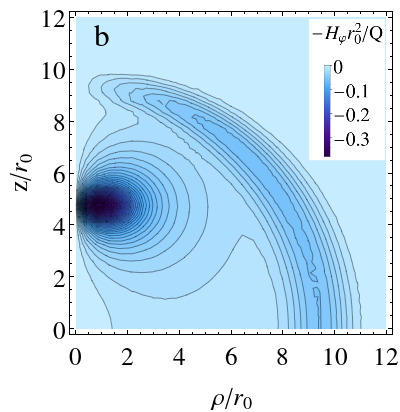}
\caption{Magnetic field $H_\phi$ distribution (in the plane passing through the Z-axis) from the FDTD simulation (a) in comparison with the theoretical result (b) for $v=0.5 c$ at the instant, $t = 10 t_0$. } \label{fig5}
\end{figure}

\begin{figure} [!ht]
\centering \includegraphics[width=0.43 \linewidth]{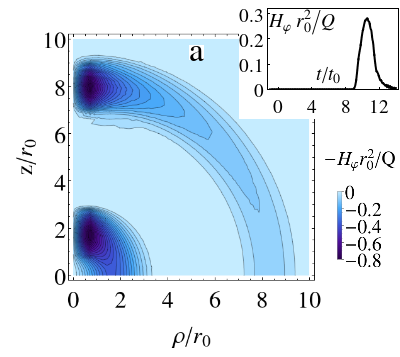}
\includegraphics[width=0.32 \linewidth]{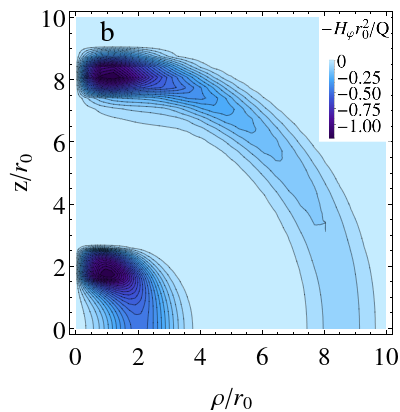}
\caption{The same as in Fig. \ref{fig5}, but  for $v=0.95\, c$ at $t = 2.7 t_0$ (bottom) and $t = 9.1 t_0 $ (top). The inset shows the magnetic field temporal profile at the distance of $10 r_0$ along the direction in which a radiated field is maximum.} \label{fig6}
\end{figure}

\begin{figure} [!ht]
\centering \includegraphics[width=0.43 \linewidth]{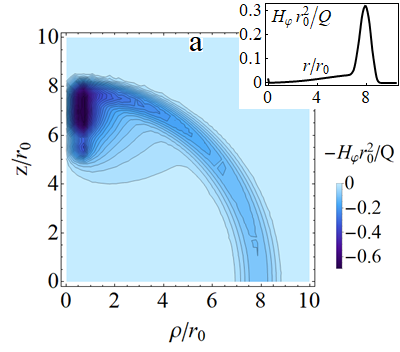} \includegraphics[width=0.43 \linewidth]{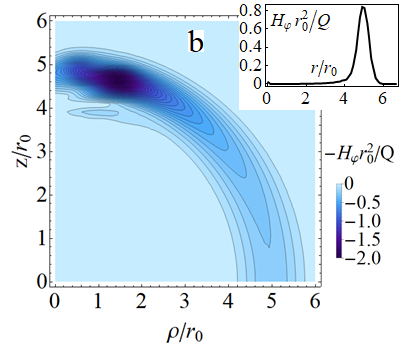}
\caption{(a) Magnetic field ($H_\phi$) distribution from the FDTD simulation at $t=9 t_0$ for the case of the electrons distributed over velocities and escaping in the normal direction. (b) Magnetic field distribution from the FDTD simulation at  the instant $t=5.6 t_0$ for the case of the electrons escaping with different velocities inside a cone with the open angle of $15^\circ$. The insets show the magnetic pulse spatial profiles along the direction corresponding to the angle of 45$^o$.} \label{fig7}
\end{figure}

Like the analytical theory, the performed simulations demonstrate formation and propagation of the half-cycle terahertz pulse. The field distribution is in a good agreement for its parameters with the theoretical model of Sec. \ref{secB} as clearly can be seen from the comparison of the density plots for $H_\varphi$ presented in Fig. \ref{fig5} and Fig. \ref{fig6}. As it should be, somewhat smoother magnetic field distribution along z-direction from the simulation results is due to the use of the Gaussian temporal charge beam profile instead of the rectangular one used in the theoretical model.

We also verified that the unipolarity of the THz pulse is conserved if the electrons of the radiating bunch have a given energy distribution, so that the faster electrons can overtake the slower ones (see Fig. \ref{fig7}a). We have simplistically chosen groups of escaping electrons with $v=0.7\,c$ and $v=0.99\, c$ distributed in accordance with the Boltzmann distribution, $\propto e^{-{\cal E}_e/\Delta T_f}$ and $\Delta T_f =4$ MeV that corresponds to the effective velocity spread $\Delta \hat{v} \simeq 0.76\, c$.

Since generation of the THz radiation by electrons moving in a certain angle range looks more realistic, we also performed  corresponding simulation.  The results presented in Fig. \ref{fig7}b for the three groups of electrons escaping inside a cone with the open angle of $15^\circ$. The electrons with the slowest velocity, $ v = 0.95 \,c$ had a Gaussian spatial distribution with characteristic scale $r_0$ and uniform distribution over the angles, $\theta$, raged from $\theta=0^\circ$ to $\theta=15^\circ$.  Other, more energetic electrons ($ v = 0.99 c$ and $v = 0.999 c$), had the same spatial distribution but escaped under zero angle. All three groups of electrons were distributed according to the Boltzmann distribution with $\Delta T_f=4$ MeV. The
simulation has been performed in the box $-100\,\mu$m$<x<100\,\mu$m, $-100\,\mu$m$<y<100\,\mu$m, $-67\,\mu$m$<z<133\,\mu$m, where a target is placed at $z<0$. The electron bunch had the sizes $c t_0 = 5\, \mu$m and $r_0 = 10\, \mu$m. As for the previously considered models of the electron source, generation of the half-cycle terahertz pulse is also clearly seen in Fig. \ref{fig7}b.

\section{Discussion and summary}

The THz wave field amplitude is proportional to the total charge, $Q$, of escaping high-energy electrons, making up only a small fraction of entire laser-heated electrons. These electrons must have enough energy to overcome the potential barrier, $ \Phi_m $, to leave the target. The characteristic value of this potential barrier at the target-vacuum interface is $e \Phi_m = -2 T_h \ln[r_0/(\lambda_{De} \sqrt{2})]$, where $T_h$ is the temperature of hot electrons with the density $n_h$. Correspondingly, the escaping electron density, $n_f$, can be estimated as $n_f \simeq n_h \exp{(e \Phi_m/T_h)}$, i.e. the total charge reads $Q =e n_f c t_0 \pi r_0^2 = T_h c t_0 /(2 e)$. The hot electron temperature standardly follows to the ponderomotive scaling, $T_h \simeq m c^2 (\sqrt{1 + a_0^2/2} -1)$, that leads to $T_h\approx 0.7\times m c^2 a_0$ for relativistically intense laser pulse, where $a_0$ is the dimensionless laser field amplitude, $a_0= 0.85 \sqrt{I [10^{18}\mbox{W/cm}^2]/\lambda_0[\mu\mbox{m}]}$ ($I$ is the laser pulse intensity and $\lambda_0$ is the laser wavelength). Finally, the total charge of the escaping electrons depends only on the amplitude and duration of the laser pulse, $Q = 0.35\times e a_0 c t_0 /r_e$, where $r_e = e^2/m c^2$ is the classical electron radius.

The total bunch charge roughly estimates the total radiated THz energy, ${\cal E}_R$, as ${\cal E}_R \sim Q^2/c t_0 \sim 0.1\times m c^2 a_0^2 c t_0 /r_e$, as well as the conversion efficiency, $\eta$, of the laser pulse energy ${\cal E}_L = m c^2 a_0^2 c t_0 R_0^2 \pi/(2\lambda^2 r_e)$ into the radiation energy, $\eta  = {\cal E}_R/{\cal E}_L \sim \ 0.08 \lambda^2/R_0^2$.
Here $R_0$ is the laser focal spot radius, which may differ from the electron emitting spot radius, $R_0 < r_0$.  For the given laser energy a tight focusing is favorable for THz radiation production. For example, when focusing a laser beam into a $4 \lambda$ spot the conversion efficiency reaches 2 \%.
A ten joule laser pulse of the 100 fs duration (100 TW) focused into (2-3) $\lambda_0$ focal spot produces broadband (up to 10 THz) $\sim100$ mJ unipolar THz pulse with the field amplitude $\sim 10^{10}$ V/m at the distance of 1 mm from a target, that is close to the record value published to date  \cite{liao20,Tzortzakis}. A significant increase in the intensity of the terahertz pulse can be expected when the femtosecond laser pulse is focused into the diffraction limit.

The presented theory analytically describes production of unique half-cycle THz pulses from a back side of laser irradiated foil target. This requires the target to be thin, of micron size thickness, and have large transverse size, $\gtrsim 1$ cm, to suppress contribution of the diffraction radiation \cite{tilborg04,root07}. A controlled preplasma on the irradiated side of the target could make it possible to achieve the maximum current of the electron current emitted from behind and, hence, to maximize the yield of THz radiation.

Unlike the previous ones, the developed theory describes the structure of the generated THz fields in the entire vacuum region, from the near to the far zone. As has been demonstrated,  for ultrarelativistic electrons the far-zone approximation is applicable at very long distances, where the emitted pulse is already weakened. The analytical theory and the long scale FDTD simulations open the way to planning an experiment to detect superstrong terahertz fields near the target surface, e.g., using laser-produced charged particles as an invaluable tool for the probing of the electric and magnetic fields \cite{Quinn}.

When this paper has already been written we were aware of experiments on THz pulses generated when femtosecond laser pulse irradiates thin foil with specially designed preplasma \cite{savelev}. The results presented there indicate quasi-half-cycle nature of the measured pulses.

In summary, the results reported clearly demonstrate that strong THz emission generated through the transition radiation by laser-produced high-energy electrons from a solid target occurs in the form of unique half-cycle pulses. It is highly probable that, taking into account the developed theory and the performed simulations, the terahertz radiation observed in a number of experiments, e.g. Refs. \cite{liao20}, should be interpreted as generation of the unipolar THz pulses. Direct experimental confirmation of such novel view on the nature of laser triggered terahertz emission would be of great interest. A possible approach could be the using of electron or proton radiography. As a final note, we emphasize, that the theory proposed could be also applied to the quantitative description of the transient surface fields, since it may advance the previously considered 2D approach \cite{pre20}.

This research was supported by Ministry of Science and Higher Education of the Russian Federation (Agreement No 075-15-2021-1361).



\end{document}